# The use of cloud technologies when studying geography by higher school students

Olga V. Bondarenko[1][0000-0003-2356-2674], Olena V. Pakhomova[1][0000-0001-5399-8116]
and Vladimir I. Zaselskiy[2]

[1] Kryvyi Rih State Pedagogical University, 54, Gagarina Ave., Kryvyi Rih, 50086, Ukraine
`bondarenko.olga@kdpu.edu.ua, helenpah@gmail.com`
[2] Kryvyi Rih Metallurgical Institute of the National Metallurgical Academy of Ukraine,
5, Stepana Tilhy Str., Kryvyi Rih, 50006, Ukraine
`zaselskiy52@mail.ru`

**Abstract.** The article is devoted to the topical issue of the cloud technologies implementation in educational process in general and when studying geography, in particular. The authors offer a selection of online services which can contribute to the effective acquisition of geographical knowledge in higher school. The publication describes such cloud technologies as Gapminder, DESA, Datawrapper.de, Time.Graphics, HP Reveal, MOZAIK education, Settera Online, Click-that-hood, Canva, Paint Instant. It is also made some theoretical generalization of their economic, technical, technological, didactic advantages and disadvantages. Visual examples of application are provided in the article. The authors make notice that in the long run the technologies under study should become a valuable educational tool of creation virtual information and education environments connected into common national, and then global, educational space.

## 1 Introduction

### 1.1 The problem statement

The essential characteristic of nowadays society is the existence of the universal information space based on global computer networks and information technologies. So it has become vital for an educated person to be highly competent in working with large masses of information. This determines the need for the integrated use of Internet information opportunities in education.

The modern educational policy strives to train a highly skilled professional who can be mobile and flexible professionally in the information society, can easily navigate in the global information space, and, what is more important, be capable of effective self-education throughout the life time. Everything mentioned above is the key to the success in the fast-moving world [3].

The training of future teachers in general and teachers of geography, in particular, is no exception. The higher educational institutions try to fulfill the social demand to train competitive pedagogical staff linking it with the further search for ways to improve the quality of education and the level of competence. One of the ways to do these is to implement cloud technologies in the educational process.

Cloud technologies are the fundamentally new services that allow you to remotely use the tools of data processing and storage, provide Internet users with access to computer resources of the server and use software as an online service [42].

## 1.2    Theoretical background

In the scientific literature there are widely represented the general scientific aspects of the problem under study, as well as those relating to specific teaching methods. In particular, scientists examine the essence of the concept, types of educational clouds, forms and necessary components of cloud technologies (Svetlana H. Lytvynova [14; 15]); theoretical foundations, possibilities and methods of using cloud technologies in the educational process are studied by Oksana S. Makovoz [16], Oksana M. Markova [17], Yevhenii O. Modlo [23], Vadym S. Nazarenko [26], Pavlo P. Nechypurenko [22], Tetiana S. Perederii [16], Maryna V. Rassovytska [30], Serhii O. Semerikov [18; 33], Vladimir N. Soloviev [35], Andrii M. Striuk [30; 33], Mariya P. Shyshkina [32], Illia O. Teplytskyi [34] researches the benefits and prospects of using cloud technologies in the educational process of a modern school.

The use of cloud technologies in computer science classes is studied by Anna I. Gazeikina [10], Olha V. Korotun [13], Oksana M. Markova [19], Susana N. Seytveliyeva [36], Viktoriia G. Shevchenko [37], Mariia V. Stupina [40]; in maths classes by Georgii A. Aleksanian [1], Maiia V. Popel [29], Tatiana A. Vakalyuk [42]; in history classes by Olena V. Burlaka [7], in languages classes by Irina A. Belysheva [3], Olena V. Pakhomova [28], Lucie Renard [31], Nataliia V. Skrynnik [38]; in physics classes by Maksym V. Khomutenko [12], Oleksandr V. Merzlykin [15]; for the formation of self-educational competence of future specialists (Tetiana V. Voloshyna [43]) and etc.

The use of cloud technologies during the study of geography is mostly represented in the works of foreign scientists, in particular Anna Badia i Perpinyà, Montserrat Pallarès Barberà and Joan Carles Llurdés i Coit [2]; Richard G. Boehm and Cheryl A. Frazier [8], Andrew J. Milson [21], Jacqui Murray [25], Tiani Page and Beverly J. Christian [27], Michael Zimmer [44].

The analysis of works on methods of geography teaching suggests that the most commonly used in secondary schools are general-purpose cloud technologies such as Google Apps, LearningApps.org, scribing technologies, blogs, online puzzle-maker generators, ribbons, tests, etc. They help develop geography knowledge and provide educational communication while studying.

Higher schools often use GIS technologies (Google Earth, Google Maps, DataGraf, Microsoft Map, Map Info, ArcGIS) that allow students to work with cartographic material [24].

At the same time, it should be noted that both approaches may be considered one-sided, since only complex systematic use of cloud technologies (both general purpose and GIS technologies), provided that the pedagogically balanced combination of traditional and innovative technologies of classroom, distance and mobile learning will contribute to effective achievement of didactic goals and thorough study of geography. Moreover, the scientists and teachers fail to notice the advantages of those cloud technologies that help work with accounting and statistical data that are widely included in study of economic and social geography. The cloud technologies mentioned above aid to memorize the large amount of geographical nomenclature and the creation of not only maps, but also dynamic charts, logical reference schemes, etc. In this article, we consider the selection of such cloud technologies that can be helpful not only for students of geography major, but also for everyone who deal with geography study in higher school.

### 1.3    The objective of the article

The purpose of the proposed publication is to characterize the content of some cloud technologies that should be used in geography students training in higher school.

## 2    Presenting the main material

The first cloud technology to be considered is *Gapminder*. We cannot deny the leading role of maps while geography studying, but we want to draw attention to the resource Gapminder. It is developed by the Gapminder Swedish fund, with Hans Rosling as a co-founder, that collaborates with educators all around the world. Using the Trendalyzer software Gapminder visualizes statistics in the form of interactive charts. This cloud environment enables a student to analyze databases provided by the following organizations: World Bank, FAO, International Labor Organization, World Health Organization, UNAIDS, IARC, etc. The statistics can be optionally illustrated from 1800 to 2018 at the global, regional or local (individual country) levels. With the help of the time tape (as well as maps, ratings, sex-age pyramids, presentations, videos), one can trace the nature of the dynamics of demographic phenomena and processes, carry out a comparative analysis of various quantitative or qualitative demographic indicators (population, child mortality, urbanization, welfare of the countries, education of the population, poverty and many others), Fig. 1.

Carmine Gallo rightly calls well-known statistician Hans Gösta Rosling a star among the TEDers, since his presentation at the TED conference in 2006 was unrivaled and "became an online "viral sensation" [9]. We are convinced that the introduction into accounting and statistical data while studying economic and social geography (especially the geography of the population) is bound to be started with Gapminder and fragments of the video collection "Do not panic. The truth about population" by Hans Rosling. Being presented by the distinguished TEDer, "grey" statistics which students do not really enjoy turns into a clear and distinct language of convincing facts that create a coherent demographic picture of the world.

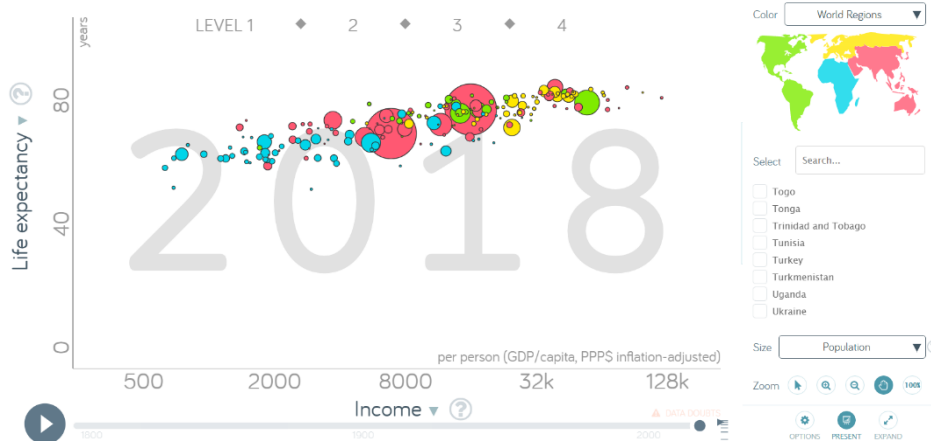

**Fig. 1.** Interactive graph "The population of the world", built in Gapminder

The second cloud technology under consideration is DESA Technology (Department of Economic and Social Affairs) / Population division, United Nations). This resource makes it possible to build and analyze randomly demographic profiles and probabilistic forecasts that reflect key demographic indicators from 1950 to 2017 for countries or different world regions, Fig. 2.

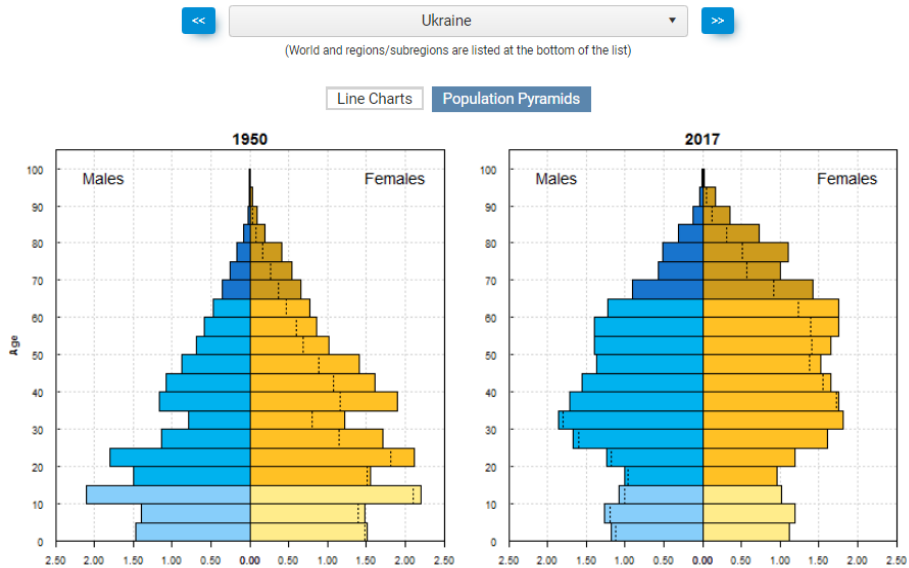

**Fig. 2.** Gender-age pyramids of the population of Ukraine, DESA

The third cloud technology is *Datawrapper.de.* This service contains an online designer and can create an interactive table or map in four steps: download, data validation, visualization and publication.

The benefits of this cloud technology are: the simplicity and speed of a final product creation; the access to the finished charts and maps folder and a wide set of templates; the import data from Excel, CSV, PDF; the export data (PNG, PDF), and the possibility to integrate data into the environment of any site; the facility to create a new personal account, etc. However, full-customized version with custom maps, print-export and CMS integration is possible only on subscription terms. Nevertheless, the technology successfully handles with the simple geographic tasks that require data visualization.

The next resource is *Time.Graphics.* This cloud technology was developed by Yevgeny Mustafin in 2017. It is aimed to visualize the chronology of events in the form of a timeline by adding photos, audio and video materials. The resource can be used while studying any discipline (history, biology, literature, foreign language). When studying geography, it is helpful to illustrate the stages of formation of the world, region, or country political map, the process of getting country independence, the periodization of the era of Great geographical discoveries, the history of the economy branches development, etc.

The main advantages of this technology are: the timeline speed making, the intuitive interface, the wide range of backgrounds and the colors of the presentation formats of the tape, the different tools to import data (Google ArtProject applications, Google Docs, Google Maps, Google Drive, Google Sheets, Google Slides, YouTube, etc.) and export the final product.

One more cloud technology worth mentioning is *HP Reveal (Aurasma).* It is a cloud-based augmented technology [41] created in Cambridge in 2011. This platform is free of charge and easy to use. The key point of HP Reveal technology is the uploading of an aura that is a visual trigger with overlaid multimedia content. By using a camera on a smartphone or tablet, the technology recognizes real images and displays overlaid animation, video, 3D model, or webpage (Fig. 3).

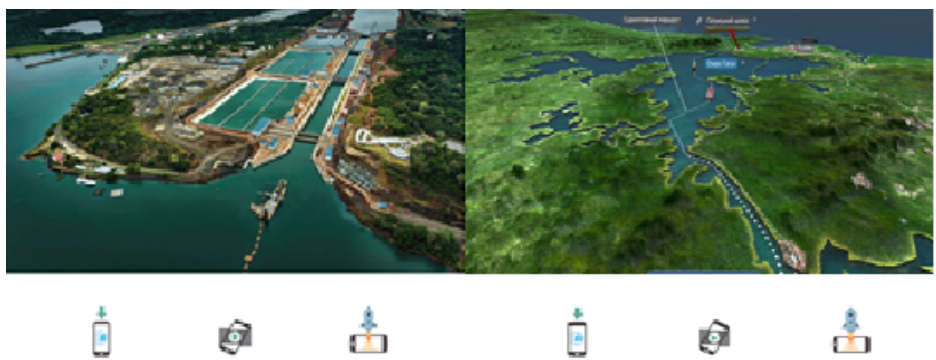

**Fig. 3.** Aura "Panama Canal", created by HP Reveal

The main advantage of this technology is the expanded didactic capacity of any printed publications, school textbooks, manuals, etc., achieved by visualizing its educational content. The main obstacle in its implementation may be as that a personal mobile device be equipped with the Aurasma App (HP Reveal), to a certain extent restricting the access to aura. Aurasma can be used while studying many higher school disciplines, but let us dwell on geography studies.

All the concepts that students master while studying geography can be divided into concrete and abstract ones with their own specifics. Thus, for the disclosure of specific concepts (river, mountain, enterprise, etc.), the inductive method is traditionally used, and for the comprehension of abstract concepts (climate, weather, historical and geographical region, specialization or concentration of production) is more often applied deductive one. Since concrete concepts are often available for direct perception, students encounter less difficulty with them than with abstract concepts. So 3D models, videos and interactive images are ultimate teaching tools, because the vivid outlook on some geographical phenomena or process contributes better to the conscious comprehension of abstract complicated concepts.

The next *MOZAIK education.* This is a service software of the Hungarian company specializing in educational interactive software for almost three decades. The resource is available in 30 languages and has both free and paid content.

Taking into accounts the specifics of geography studies, a media library that is a cloud storage and contains digital lessons, 3D scenes, videos, images, audio, task letters, tools and games offered for primary, secondary and higher schools can be of great value and importance.

The advantages of the considered service include: the high resolution of the video and 3D models (Fig. 4), which demonstrate components of the geographical phenomenon or process, their integral representation in the form of a video animation; review normal, anaglyph or stereoscopic modes; scaling the user interface of a 3D player; tools to work with the legend, separate layers of the image, additional information and facilities to make virtual trips.

In addition, the "Tools and Games" category includes three-dimensional maps and tasks, interactive maps, three-dimensional images of the Earth, a collection of aerial photographs, a constructor for quick and easy creation of diagrams using built-in templates, interactive test editor, etc.

However, free viewing of 3D content is limited to 5 units per week. Premium mozaWeb subscription is needed for full access to the library contents, search and playback of 3D models, videos, audio materials or images. The subscription enables further e-learning and working with interactive tutorials.

The next point in cloud technology collection is *Settera Online.* This popular service is adapted in 32 world languages and is intended for studying the geographical nomenclature. Swedish programmer Marianna Wartoft developed Settera Online in 1998. The service exists both in the computer version and in the form of mobile applications for iOS, Android and iPad.

This cloud application can replace the traditional method of learning numerous geographical names into an interactive one [4]. Before now, a student showed geographical positions and locations on the map in the classroom to demonstrate his knowledge. Now students have to find the geographical position of the country or capital, recognize its outline in a limited time, and then send the answer to a teacher for a check. If the answer is error-free and done on time, the map will paint white, assuming one error – yellow, and two errors – red. So, in order to cope with the task in the set time, students have to perform the task online at least ten times. The number of practicing promotes better memorization of the geographical names eventually.

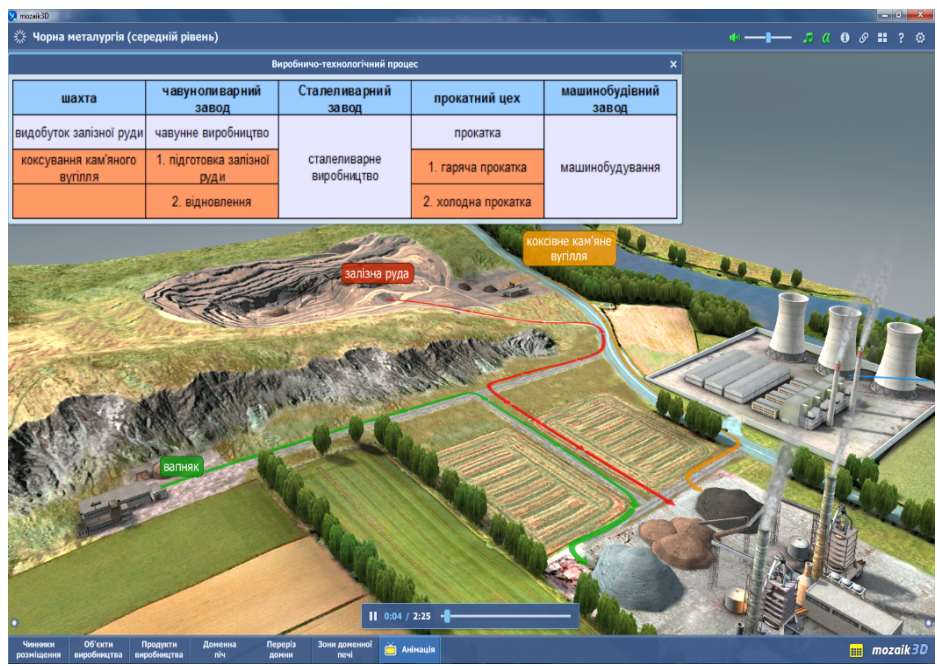

**Fig. 4.** Ferrous metallurgy, 3D scene MOZAIK education

The advantages of such method of nomenclature study are obvious. They are: an individual pace of a task performance; objectivity of assessment; developed mapping skills; rational use of classroom time.

*Click-that-hood* is a cloud technology for studying and verifying the names of units of administrative and territorial division of countries (states, regions, provinces, lands, prefectures, voivodships, etc.), a technique similar to that of Settera Online.

A very popular present day form of data visualization is the infographic, which does not take much time to view and provides the reception of large volumes of information in an accessible and comprehensible form. The leading Ukrainian resource that supplies accounting and statistical data in the form of infographic directories is *BusinessViews* [39]. Unfortunately, the authors cannot propose a free cloud-based technology that provides geographic knowledge in the same way as BusinessViews does. However, the Canva cloud technology can serve as an alternative.

*Canva* is a graphical design software that visualizes information in the form of a presentation, a graphic, a map, a booklet, a postcard, etc. The benefits of this service are free registration or access through Google and Facebook; user-friendly interface; wide selection of ready-made templates and filters; access to a folder of photos, illustrations, fonts, styles, etc.

In this article we do not characterize special cartographic editors that can be used for studying geography, since a separate publication will be dedicated to the consideration of this issue. However, some of the cloud technologies mentioned above help create maps, including Gapminder, Datawrapper.de, Canva. Google *Instant Paint* Image Editor has also a potential for creating geographic mappings. The method of work in the environment of this service is similar to Paint. Although it is not possible to create proper geographic maps, however, charts or logical reference schemes reviling geographic knowledge of a student can be quite informative (Fig. 5).

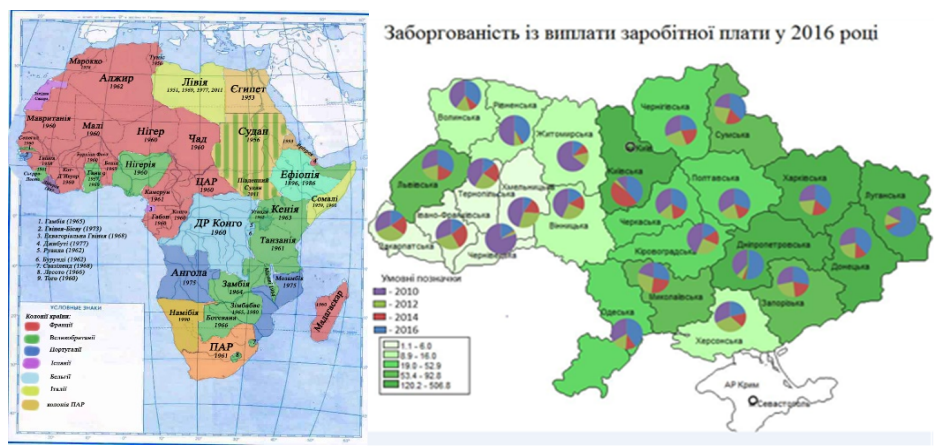

**Fig. 5.** Cards created in Paint Instant

Unfortunately, the format of one article does not allow to consider all available cloud technologies that can be used for studying geography. However, the implementation of the services described by the authors can significantly expand the methodical arsenal of teachers and higher students and make the educational process more efficient.

Today, it is impossible to format a student coherent geographic picture of the world in a higher school only by traditional didactic means. Geography is a science to be studied when traveling and instantly perceiving the object of study [5; 6]. That is why cloud technologies make it possible to visualize geography learning in a virtual environment and even substantially increase the perception of geographic processes not available in direct observation. What previously seemed to be difficult, and sometimes inconceivable, is quite possible at the moment (for example, to look at the volcano's burning stove or blast furnace, to observe the formation of the Earth or the work of a nuclear power plant, etc.). What is more, the generation of students who cannot live without gadgets, should learn how to use them rationally rather than be deprived of them.

The analysis of the cloud technologies observed above and their implementation in educational purposes internationally proves their long-term benefits due to the number of advantages. We fall into line with Anna I. Gazeikina and Alevtina S. Kuvina [10], and we consider their pluses can be summarized in the following groups:

— economic benefits (free of charge or privileged access to the majority of services; less number of auditorium and equipment required for training; reduction of the number of personnel required for maintenance);
— technical benefits (minimum hardware requirements on condition of access to the Internet, lack of technical support for the work of the platform and configuration);
— technological benefits (high quality and intuitive interface of the majority of cloud technologies; personal data protection and delimitation of common information access; rapid integration of created products into the educational process; no attachment of the service user to the territorial location);
— didactic benefits (a wide range of online tools and services that can be implemented with a different didactic purpose and at different stages of the classroom; the variety of interactions; simplification of information creation, accumulation and exchange; the expansion of out of classroom training opportunities; the increase in the academic performance and students' internal motivation, etc.).

However, the implementation of cloud technologies in the educational process has also some disadvantages. They can be the following: limited access to the services or subscription requirement; limited functions of online software in comparison with the local one; the absence of Ukrainian cloud service providers; high quality requirements for enforced path; the dependence of non-stop operation and important data storage on the service provider; possible errors and leakage of information with user increase; lack of legislative framework for the use of cloud technologies; low level of computerization of Ukrainian institutions of higher education; insufficient development of theoretical and methodological principles of cloud technologies implementation in the educational process; the unwillingness of some teachers to combine traditional and innovative educational technologies, as the implementation of the latter requires additional efforts; underestimation of the importance of cloud technologies in the professional development of teachers and students [4; 10; 16; 26].

We want to emphasize that cloud technologies should gradually become a thorough didactic means, enabling educational institutions to create their own virtual information and education environments, with the prospect of integrating into a common national, and then global, information space.

## 3    Conclusion

1. Nowadays decision-making is often based on the information of various Internet sources [11], so an educated person is required to be highly competent in dealing with large amounts of information. Due to the relevance of this problem for the IT society, one cannot overestimate the didactic prospects of cloud technologies, as they contribute to: efficiency of handling with the students' real life problem situations, which can be sorted out with digital devices and gadgets [26]; mastering the skills to find, systematize, analyze a large amount of necessary information; the reasonable use of cloud technologies, the skills to assess the benefits and risks of cloud technologies for self-development, environment or society.
2. We see the prospects of further scientific research in the study of the practical cloud technologies implementation while the geography study in higher school.